# Theory of magnetization precession induced by a picosecond strain pulse in ferromagnetic semiconductor (Ga,Mn)As

T. L. Linnik[1], A. V. Scherbakov[2], D. R. Yakovlev[2,3], X. Liu[4], J. K. Furdyna[4], and M. Bayer[2,3]

[1] *Department of Theoretical Physics, V. E. Lashkaryov Institute of Semiconductor Physics, National Academy of Sciences of Ukraine, 03028 Kyiv, Ukraine*

[2]*Ioffe Physical-Technical Institute, Russian Academy of Sciences, 194021 St. Petersburg, Russia*

[3]*Experimentelle Physik 2, Technische Universität Dortmund, D-44227 Dortmund, Germany*

[4]*Department of Physics, University of Notre Dame, Notre Dame, Indiana 46556, USA*

ABSTRACT

A theoretical model of the coherent precession of magnetization excited by a picosecond acoustic pulse in a ferromagnetic semiconductor layer of (Ga,Mn)As is developed. The short strain pulse injected into the ferromagnetic layer modifies the magnetocrystalline anisotropy resulting in a tilt of the equilibrium orientation of magnetization and subsequent magnetization precession. We derive a quantitative model of this effect using the Landau-Lifshitz equation for the magnetization that is precessing in the time-dependent effective magnetic field. After developing the general formalism, we then provide a numerical analysis for a certain structure and two typical experimental geometries in which an external magnetic field is applied either along the hard or the easy magnetization axis. As a result we identify three main factors, which determine the precession amplitude: the magnetocrystalline anisotropy of the ferromagnetic layer, its thickness, and the strain pulse parameters.

# 1. INTRODUCTION

Ultrafast control of magnetic order is one of the key problems of modern magnetism. The performance of magnetic storage devices, which lags much behind their exponentially increasing capacity, is a bottleneck of current electronics. During the last decade various concepts to manipulate magnetization on a short time scale utilizing picosecond magnetic field pulses [1,2] or femtosecond optical excitation [3] have been explored for magnetic materials. In materials with strong magnetocrystalline anisotropy (MCA) acoustic pulses may be also an effective tool to manipulate magnetization on ultrashort time scales [4,5]. The methods of picosecond laser ultrasonics allow the generation of ultrashort strain pulses in solids [6]. These strain pulses have picosecond duration and amplitude up to $10^{-3}$. They have a fast and local impact, which may lead to a considerable response of the material's magnetization, whose magnetic properties are sensitive to strain.

Ferromagnetic semiconductors (FMSs), like (Ga,Mn)As, belong to the class of ferromagnets with strong MCA due to the hole-mediated origin of ferromagnetism [7,8]. In FMS epitaxial layers mainly strain determines the directions of the easy magnetization axes. The compressive (tensile) epitaxial strain from lattice mismatch between buffer and FMS layers results in in-plane (out-of-plane) orientation of the easy axes of magnetization for a wide range of FMS parameters [9,10,11]. Several ways to control the magnetization in FMS by strain have been developed recently: (i) the desired direction of the easy magnetization axis may be achieved by adjusting the composition of a buffer layer during growth [9]; (ii) after-growth patterning allows directing the in-plane magnetization [12]; and (iii) in layered multiferroic structures with the FMS layer grown on piezoelectric material an electric field applied to the piezoelectric layer governs the in-plane unidirectional strain and allows manipulation of the magnetization direction [13-15].

Until very recently the strain-control of magnetization in FMSs has remained static. First time-resolved experiments with strain pulses in FMS epitaxial layers were reported by *Thevenard et al*. [16] and *Scherbakov et al*. [5] in 2010. The studies in Ref. [16] focused on elasto-optical effects induced by a strain pulse propagating in a magnetized FMS layer, while the effect of the strain pulse on the magnetization and the strain-induced temporal evolution of magnetization were studied in Ref. [5]. It was demonstrated in a magnetic field normal to the ferromagnetic layer that the strain pulse induces a pronounced tilt of magnetization out of its equilibrium orientation and subsequently coherent magnetization precession. In Ref. [5], for describing the experimental results the authors considered the simplest model of magnetocrystalline anisotropy of a FMS layer. The proposed model cannot explain a number of effects observed in the later experiments, such as strain pulse induced magnetization precession also for in-plane magnetic fields and even without external field [17]. This observation has stimulated the present theoretical studies, which are aimed at carrying out a comprehensive analysis of the effect of strain pulses on the magnetization in ferromagnetic (Ga,Mn)As. The main goal is to examine how the amplitude of the strain-pulse-induced precession depends on the parameters of the FMS structure, the magnetic field strength and direction and the parameters of the strain pulse. We examine the cases of magnetic field direction normal to the ferromagnetic layer as in Ref. [5] and also parallel to it as well as without magnetic field. The underlying anisotropy parameters of the FMS structure have been obtained using the microscopic model for hole-mediated ferromagnetism proposed by *Dietl et al.* [18].

The paper is organized as follows. In Section 2 we briefly describe the considered experiments with picosecond strain pulses hitting FMS layers, introduce the parameters of the strain pulse and qualitatively discuss the effect of the strain pulse on the magnetization. Section 3 describes the formalism, which is used later to calculate quantitatively the effect of the strain pulse. In Section 4 we present the results of numerical calculations for a particular FMS structure subject to two different orientations of external magnetic field. Finally, we summarize and conclude the obtained results and discuss the perspectives for controlling magnetization by picosecond acoustics.

## 2. EXPERIMENTS WITH PICOSECOND STRAIN PULSES
## IN EPITAXIAL (Ga,Mn)As LAYERS

Figure 1(a) shows the schematic of experiments with picosecond strain pulses applied to a FMS layer. The sample consists of a single $Ga_{1-x_{Mn}}Mn_{x_{Mn}}As$ FMS layer grown on a semi-insulating GaAs substrate [5]. The typical content of Mn atoms in the FMS layer is $x_{Mn} = 0.01 \div 0.1$. A thin metal film deposited on the back side of the GaAs substrates serves as optoelastic transducer, which rapidly expands due to the heating under femtosecond laser excitation [6]. Figure 1(b) demonstrates the bipolar strain pulse $\delta\varepsilon_{zz}(t)$ injected into the substrate as result of the thermal expansion of the metal film [19,20]. Pulse duration $\tau$ and amplitude $\varepsilon_{zz}^{max}$ depend on the transducer material and the parameters of optical excitation, and have typical values of ~10 ps and ~$10^{-4} \div 10^{-3}$, respectively. It is important to note, that in high symmetry GaAs substrates (typically (001) oriented) the strain pulse contains only longitudinal components for lattice distortions along the propagation direction perpendicular to the substrate interface. At liquid helium temperatures such a strain pulse propagates through GaAs over millimeter distances without scattering [21].

In order to describe the response of the magnetization **M** of the FMS layer on the strain pulse we use the standard Landau-Lifshitz approach in which the magnetization is precessing about the time-dependent effective magnetic field **B**$_{eff}$ [22]. This effective field is the sum of the external magnetic field **B** and the intrinsic magnetic anisotropy field, which is determined by the parameters of the FMS layer. In equilibrium the magnetization **M** is parallel to **B**$_{eff}$. As an example, Fig. 2(a) shows the experimental geometry reported in Ref. [5] when **B** is applied normal to the (Ga,Mn)As layer with in-plane easy axes. In such a layer the anisotropy field holds **M** in the layer plane, while the external magnetic field turns **M** out of the layer, so that the resulting field **B**$_{eff}$ has a tilted orientation between in-plane and normal-to-it. When reaching the FMS layer, the strain pulse changes the layer properties, namely the $\varepsilon_{zz}$ static strain component, modifies the magnetic anisotropy field, and tilt**s** **B**$_{eff}$, which is then no more parallel to **M**. As a result **M** starts to precess around **B**$_{eff}$. After the strain pulse has left the FMS layer, **B**$_{eff}$ returns to its equilibrium orientation, while **M** remains at some angle relative to **B**$_{eff}$. Thus, the precession continues until relaxation drives **M** back to equilibrium [Fig. 2(b)]. In the Landau-Lifshitz approach value and direction of **B**$_{eff}$ are determined by the free energy density [23]. The free energy density includes magneto-elastic terms, which provide the direct relation between the strain components and the orientation of **B**$_{eff}$. Thus, one can model the response of **B**$_{eff}$ and the magnetization on the strain pulse, as shown in the next Section.

## 3. MAGNETIZATION PRECESSION INDUCED BY A STRAIN PULSE

In our theoretical analysis we consider a thin FMS (Ga,Mn)As layer with a typical Mn ion content that is epitaxially strained, at liquid helium temperatures. Figure 2(a) shows the assumed coordinate system, in which the *x* and *y* axes lie in the layer plane along the [100] and [010] crystallographic directions, respectively, and the *z*-axis is perpendicular to the layer growth direction, which is the [001] crystallographic direction. Far below the Curie temperature the magnetization of the FMS layer is close to the saturation value $M_0 = g\mu_B N_{Mn} S_{max}$, where *g*=2 is the Mn Lande factor, $\mu_B$ is the Bohr magneton, $S_{max} = 5/2$ is the maximal total spin of the Mn atom and $N_{Mn} = 4x_{Mn}/a_0^3$ is the concentration of Mn atoms ($a_0$ is the lattice constant). Assuming that the perturbation induced by the strain pulse is weak and does not affect the absolute value of $M$, and neglecting also damping we may use the Landau-Lifshitz equation to describe the dynamics of magnetization in the time-dependent effective field $\mathbf{B}_{eff}(t)$ [22]:

$$\frac{d\mathbf{m}}{dt} = -\gamma \cdot \mathbf{m} \times \mathbf{B}_{eff}(\mathbf{m},t), \qquad \mathbf{B}_{eff}(\mathbf{m},t) = -\nabla_{\mathbf{m}} F_M(\mathbf{m},t), \qquad (1)$$

where $\mathbf{m} = \mathbf{M}/M_0$ is the normalized magnetization and $\gamma = g\mu_B/\hbar$ is the gyromagnetic ratio. The effective field $\mathbf{B}_{eff}$ acting on $\mathbf{m}$ is determined by the gradient of the normalized free energy density of the FMS layer $F_M = F/M_0$.

Generally, the free energy density $F_M$ consists of isotropic and anisotropic parts. The isotropic part does not depend on the direction of $\mathbf{m}$ and does not contribute to the vector product in Eq. (1). Therefore, we have to consider only the anisotropic part of $F_M$, which includes the Zeeman term, the demagnetization energy, and the MCA terms related to the crystal symmetry. In a thin (Ga,Mn)As layer grown by low-temperature molecular beam epitaxy the cubic symmetry is tetragonally distorted by the epitaxial strain originating from the lattice mismatch between the buffer and the (Ga,Mn)As layers. Most of experiments also indicate the presence of an in-plane uniaxial anisotropy in the (Ga,Mn)As films [10,11,24]. The origin of this anisotropy is still under discussion, but phenomenologically it can be modeled by a weak shear strain $\varepsilon_{xy}$ [24]. Thus, we write the general expression for the anisotropic part of the free energy density of a thin cubic FMS layer distorted by strain [23,25,26] in the form:

$$\begin{aligned} F_M(\mathbf{m}) = &-(\mathbf{m}\cdot\mathbf{B}) + B_d m_z^2 + B_c(m_x^4 + m_y^4 + m_z^4) + A_{2\varepsilon}(\varepsilon_{xx} m_x^2 + \varepsilon_{yy} m_y^2 + \varepsilon_{zz} m_z^2) + \\ &+ 2A_{4\varepsilon}^{(1)}(\varepsilon_{xx} m_y^2 m_z^2 + \varepsilon_{yy} m_x^2 m_z^2 + \varepsilon_{zz} m_x^2 m_y^2) + A_{4\varepsilon}^{(2)}(\varepsilon_{xx} m_x^4 + \varepsilon_{yy} m_y^4 + \varepsilon_{zz} m_z^4) + A_{2xy}\varepsilon_{xy} m_x m_y \end{aligned} \qquad (2)$$

where $m_x$, $m_y$ and $m_z$ are the projections of $\mathbf{m}$ onto the coordinate axes and $\varepsilon_{ij}$ ($i,j=x,y,z$) are the strain components. The first term in Eq. (2) is the Zeeman energy of $\mathbf{m}$ in the external magnetic field $\mathbf{B}$, the second term is the demagnetization energy of the thin ferromagnetic film with $B_d = \mu_o M_0/2$ [27,28], and the five following terms describe the MCA of the strained cubic

FMS layer. The cubic anisotropy field $B_c$ and the magnetoelastic coefficients $A_{2\varepsilon}$, $A_{4\varepsilon}^{(1)}$, $A_{4\varepsilon}^{(2)}$, and $A_{2xy}$ are parameters of the FMS film, which depend on lattice temperature, hole concentration $p$ and Mn content $x_{Mn}$ [9-11,24,29]. The equilibrium orientation of **m** is given by the minimum of $F_M$ and depends on the balance between Zeeman, demagnetization and MCA energies.

In the unstrained FMS layer the MCA part of $F_M$ in Eq. (2) consists of the cubic term proportional to $B_c$ only. For the experimentally relevant ranges of $p$ and $x_{Mn}$ at low temperatures the value of $B_c$ may be both negative or positive [18,29]. We consider the case $B_c<0$ when the six equivalent easy magnetization axes lie along the [100], [010] and [001] crystallographic directions. This equivalence is destroyed by the static epitaxial strain with components:

$$\varepsilon_{xx} = \varepsilon_{yy} = (a_0 - a)/a, \qquad \varepsilon_{zz} = -2\varepsilon_{xx} \cdot C_{12}/C_{11}, \qquad (3)$$

where $a_0$ and $a$ are the non-distorted lattice constants of the (Ga,Mn)As and GaAs layers, respectively. $C_{11}$ and $C_{12}$ are the elastic modules of (Ga,Mn)As. As a result the in-plane [100] and [010] and the out of plane [001] orientations of **m** become nonequivalent. At low temperatures, for sufficiently high hole concentrations in-plane compressive strain $\varepsilon_{xx} = \varepsilon_{yy} < 0$ is found in (Ga,Mn)As layers grown on GaAs, leading to in-plane orientation of the easy axes [10,11,18,29]. Further, the in-plane uniaxial anisotropy determined by the last term of Eq. (2) leads to a tilt of the easy magnetization axis from the [100]/[010] crystallographic directions toward [1$\bar{1}$0]/[$\bar{1}$10] for positive $\varepsilon_{xy}$. This means that the coefficients $A_{2xy}$ and $A_{2\varepsilon}$ must be positive. The cubic magnetoelastic coefficients $A_{4\varepsilon}^{(1)}$ and $A_{4\varepsilon}^{(2)}$ are one order of magnitude smaller than $A_{2\varepsilon}$ and, consequently, do not affect the orientation of the easy magnetization axis. Finally, the demagnetization energy supports the in-plane orientation of **m**.

In the microscopic model used for the calculating the anisotropy coefficients the relation $A_{4\varepsilon}^{(1)} = A_{4\varepsilon}^{(2)}$ is fulfilled (see Appendix A) so that we will apply this approximation throughout the rest of the paper using the notation $A_{4\varepsilon}^{(1)} \equiv A_{4\varepsilon}$. Since also $\varepsilon_{xx}=\varepsilon_{yy}$ for epitaxial strain we may simplify Eq. (2) and rewrite it in spherical coordinates:

$$\begin{aligned}F_M(\theta,\varphi) = & \left[B_d + (A_{2\varepsilon} - 2A_{4\varepsilon})(\varepsilon_{zz} - \varepsilon_{xx})\right]\cos^2\theta + \\ & + \left[B_c + 2A_{4\varepsilon}(\varepsilon_{zz} - \varepsilon_{xx})\right]\cos^4\theta + \left[B_c - A_{4\varepsilon}(\varepsilon_{zz} - \varepsilon_{xx})\right]\sin^4\theta \cdot \frac{1}{4}(3+\cos 4\varphi) + \\ & + \frac{1}{2}A_{2xy}\varepsilon_{xy}\sin^2\theta\sin 2\varphi - B_x\sin\theta\cos\varphi - B_y\sin\theta\sin\varphi - B_z\cos\theta.\end{aligned} \quad (4)$$

This expression provides a direct relation between the magnetic anisotropy fields, which are typically used to describe MCA in most publications on FMS (Ga,Mn)As, and the strain components. The values $(A_{2\varepsilon} - 2A_{4\varepsilon})(\varepsilon_{zz} - \varepsilon_{xx})$, $B_c + 2A_{4\varepsilon}(\varepsilon_{zz} - \varepsilon_{xx})$, $B_c - A_{4\varepsilon}(\varepsilon_{zz} - \varepsilon_{xx})$ and $A_{2xy}\varepsilon_{xy}$ are usually defined as perpendicular uniaxial, perpendicular cubic, in-plane cubic and in-plane uniaxial anisotropy fields, respectively.

In the frame of the single-domain model with constant magnetization it is convenient to rewrite also Eq.(1) in spherical coordinates [30]:

$$\frac{\partial \varphi}{\partial t} = \frac{\gamma}{\sin\theta}\frac{\partial F_M}{\partial \theta}, \qquad \frac{\partial \theta}{\partial t} = -\frac{\gamma}{\sin\theta}\frac{\partial F_M}{\partial \varphi}. \quad (5)$$

Assuming that the changes $\delta\varphi$ and $\delta\theta$ of the angles $\varphi$ and $\theta$ induced by the strain-pulse $\delta\varepsilon_{zz}$ are small, we can write in linear approximation:

$$\begin{aligned}\frac{\partial \varphi}{\partial t} &= \frac{\gamma}{\sin\theta_o}\left[F_{\theta\theta}\delta\theta + F_{\theta\varphi}\delta\varphi + F_{\theta\varepsilon_{zz}}\delta\varepsilon_{zz}(t,z)\right], \\ \frac{\partial \theta}{\partial t} &= -\frac{\gamma}{\sin\theta_o}\left[F_{\varphi\varphi}\delta\varphi + F_{\varphi\theta}\delta\theta + F_{\varphi\varepsilon_{zz}}\delta\varepsilon_{zz}(t,z)\right]\end{aligned} \quad (6)$$

where the $F_{ij} = \dfrac{\partial^2 F_M}{\partial i \partial j}$ ($i,j = \varphi, \theta, \varepsilon_{zz}$) are calculated at equilibrium orientation $\theta_0(B), \varphi_0(B)$, corresponding to the static orientation of **m** at a given $B$.

Here we introduce the effective rates of strain-induced precession:

$$f_\varphi = \frac{\gamma}{\sin\theta_0} F_{\theta\varepsilon_{zz}} = -\gamma\cos\theta_0 \cdot (2A_{2\varepsilon} + A_{4\varepsilon}\sin^2\theta_0(\cos 4\varphi_0 - 1) + 4A_{4\varepsilon}\cos^2\theta_0),$$
$$f_\theta = -\frac{\gamma}{\sin\theta_0} F_{\varphi\varepsilon_{zz}} = -\gamma A_{4\varepsilon}\sin^3\theta_0 \sin 4\varphi_0. \tag{7}$$

The values of $f_\theta$ and $f_\varphi$ determine the amplitude and the direction of the tilt of **B**$_{\text{eff}}$ induced by $\delta\varepsilon_{zz}(t,z)$ for a specific static orientation of **m**. If both rates are zero, the strain pulse does not tilt **B**$_{\text{eff}}$ and, thus, does not induce any magnetization dynamics. One sees that if **m** lies in the layer plane, $f_\varphi = 0$. In addition, there are specific in-plane directions corresponding to the crystallographic directions [100], [010], and the diagonals, where $f_\theta = 0$, and a tilt of **B**$_{\text{eff}}$ by $\delta\varepsilon_{zz}(t,z)$ is impossible. This means that in a FMS layer with no shear strain ($\varepsilon_{xy}=0$) the strain pulse $\delta\varepsilon_{zz}(t,z)$ may induce a magnetization precession only when applying an external magnetic field, which rotates **m** out of the easy magnetization axis. However, the presence of shear strain ($\varepsilon_{xy} \neq 0$) allows launching of a magnetization precession by $\delta\varepsilon_{zz}(t,z)$, even at zero $B$. So the presence of at least one of these factors, either an external magnetic field or an in-plane shear strain, is crucially necessary to induce a magnetization precession by $\delta\varepsilon_{zz}(t,z)$.

The precession frequency $\omega_0$ is determined by the standard expression for the ferromagnetic resonance frequency and depends on the static orientation of **m** [30-32]:

$$\omega_0 = \frac{\gamma}{\sin\theta_0}\sqrt{F_{\theta\theta}F_{\varphi\varphi} - F_{\theta\varphi}^2} \tag{8}$$

It is worth to note, that Eqs. (5-8) cannot be applied, when the equilibrium **m** is parallel to the [001] axis, where a mathematical singularity appears [30,31]. However, it is easy to see that for this orientation of **m** any perturbation $\delta\varepsilon_{zz}(t,z)$ cannot turn the magnetization out of the equilibrium direction. Thus this orientation is not of our interest and we use Eqs (5-8) throughout the rest of the paper.

The developed formalism is well suited for strain pulses of arbitrary shape but we restrict the numerical calculations to spatial and temporal dependencies of $\delta\varepsilon_{zz}(t,z)$ typical for ultrafast acoustic experiments. In the (Ga,Mn)As film the strain has a complex shape compared to the one injected into the substrate, as result of interference of the incident and reflected components of the pulse. The spatial-temporal evolution of the strain pulse which propagates with the longitudinal sound velocity $v_l$ along the z-axis through the FMS layer with thickness $d$ can be modeled as [19,20]:

$$\delta\varepsilon_{zz}(t,z) = \frac{\sqrt{e}\varepsilon_{zz}^{(\max)}}{\tau}\left((t-z/v_l)\exp\left(-\frac{(t-z/v_l)^2}{2\tau^2}\right) - (t+(z-2d)/v_l)\exp\left(-\frac{(t+(z-2d)/v_l)^2}{2\tau^2}\right)\right), \quad (9)$$

where $e$ is the base of the natural logarithm. Time $t=0$ in Eq. (9) corresponds to the moment, when the center of the bipolar strain pulse reaches the GaAs/(Ga,Mn)As interface ($z=0$). The first term in Eq. (9) describes the evolution of the strain pulse propagating toward the open surface of the FMS layer and the second term describes the strain pulse reflected at the open surface with a $\pi$-phase shift and subsequently propagating back toward the substrate. The parameters of the strain pulse that we use for the further calculations are as follows: $\varepsilon_{zz}^{\max}=10^{-4}$, $\tau=7$ps, and $v_l = 5 km/s$. These values are typical for ultrafast acoustic experiments and correspond to the values reported in Ref. [5]. Here we do not take into account nonlinear effects, which modify the shape of the strain pulse during its propagation through the GaAs substrate. These effects are insignificant for the chosen strain pulse amplitude.

Figure 3(a) shows the time evolutions $\delta\varepsilon_{zz}(t,z)$ for three different positions inside the 200-nm-thick magnetic layer: $z=0$, 100 nm and 190 nm, which correspond to the GaAs/(Ga,Mn)As interface, the centre of the FMS layer, and the coordinate 10 nm before the open surface, respectively. It is clearly seen that the $\delta\varepsilon_{zz}(t,z)$ are not the same for the different coordinates. Thus, the strain-induced perturbation of $F_M$ is spatially nonuniform and Eq. (6) must be solved at each coordinate $z$ inside the FMS layer. Because of this nonuniformity of the perturbation, one also should add the exchange term to the expression for $\mathbf{B}_{\text{eff}}$ in Eq. (1) [22,25]. Basically exchange would lead to two effects. First, it gives rise to a frequency splitting of the magnon modes in a finite-width film [33]. This splitting can manifest itself by a beating due to interference of the split modes contributing to the strain-induced magnetization precession. For realistic (Ga,Mn)As parameters, however, the mentioned splitting is relatively small [34,35]. It is worth to mention also that since the exchange terms are proportional to the spatial derivatives of magnetization, proper boundary conditions must be introduced for the magnetization at the ferromagnetic film interfaces. It is known, however, that this affects the magnetization mainly in the quite thin regions near the interfaces [33]. Leaving these specific effects for further studies we proceed with the analysis of the case without exchange.

In the actual experiment probing of the magnetization at a certain coordinate $z$ is impossible. The experimental signal (i.e. the magneto-optical Kerr rotation) reflects the time evolution of the magnetization averaged over the layer thickness. Thus, we introduce the mean angles:

$$\overline{\delta\theta}(t) = \frac{1}{d}\int_0^d \delta\theta(z,t)dz, \qquad \overline{\delta\varphi}(t) = \frac{1}{d}\int_0^d \delta\varphi(z,t)dz. \tag{10}$$

Then, Eq. (6) may be rewritten for relating $\overline{\delta\varphi}(t)$ and $\overline{\delta\theta}(t)$ with the averaged strain-induced temporal perturbation, shown by the thick red line in Fig. 3(a):

$$\overline{\delta\varepsilon}(t) = \int_0^d \delta\varepsilon_{zz}(z,t)dz. \tag{11}$$

In the next section we solve Eq. (6) numerically for both the magnetization and the averaged magnetization as function of coordinate $z$.

## 4. NUMERICAL ANALYSIS OF STRAIN-INDUCED PRECESSION

We examine two characteristic orientations of the external magnetic field: perpendicular to the layer plane, $\mathbf{B}=(0,0,B)$, and in the layer plane along the [100] crystallographic direction $\mathbf{B}=(B,0,0)$. We present the results of a numerical analysis for certain parameters of the FMS layer. First, we analyze the static orientation of magnetization as function of the external magnetic field, calculate the field dependencies of the effective precession rates $f_{\varphi(\theta)}(B)$ and the precession frequency $\omega_0(B)$, and then model the time evolution of the magnetization induced by the strain pulse of chosen shape. We use the following parameters for the structure, which are typical for a thin (Ga,Mn)As layer: $d$=200 nm, $x_{Mn}$=0.045, $p$=4×10$^{20}$ cm$^{-3}$, and $\mu_0 M_0$=60 mT. The corresponding values of $B_c$= −35 mT, $A_{2\varepsilon}$=25 T, $A_{2xy}$=152 T and $A_{4\varepsilon}=0.5T$ were calculated in the frame of the Dietl model, for details see Appendix A. The calculations are limited to the case of compressive epitaxial strain: $\varepsilon_{xx}=\varepsilon_{yy}<0$; $\varepsilon_{zz}>0$ and, thus, in-plane orientation of the easy magnetization axes. The factor $2C_{12}/C_{11}$=0.89 in Eq. (3) is taken from Ref. [36]. The calculations are carried out for several values of the static strain components: $\varepsilon_{zz}$=(1÷3)×10$^{-3}$ and $\varepsilon_{xy}$=(0÷2)×10$^{-4}$. In the frame of the single domain model we assume that at zero external magnetic field **m** lies along the [100] direction if $\varepsilon_{xy}=0$ and along the easy magnetization axis that is closest to the [100] direction if $\varepsilon_{xy}>0$.

### A. Perpendicular magnetic field

An external magnetic field applied perpendicular to the FMS layer rotates the magnetization out of the layer plane toward the *z*-axis. In this case the strain pulse induces a magnetization precession even at $\varepsilon_{xy}=0$. Since also $\varepsilon_{xy}$ is typically at least one order of magnitude smaller then the epitaxial strain we first restrict our consideration to the case of zero shear strain, and thereafter numerically analyze the effect of nonzero $\varepsilon_{xy}$.

For zero shear strain ($\varepsilon_{xy}=0$) the orientation of **m** is characterized by $\varphi_0=0$ for any value of *B* and we may simplify the expression (4) for $F_M$ to:

$$F_M(\theta) = -B \cos\theta + \left[B_d + (A_{2\varepsilon} - 2A_{4\varepsilon})(\varepsilon_{zz} - \varepsilon_{xx})\right]\cos^2\theta + \\ + \left[B_c + 2A_{4\varepsilon}(\varepsilon_{zz} - \varepsilon_{xx})\right]\cos^4\theta + \left[B_c - A_{4\varepsilon}(\varepsilon_{zz} - \varepsilon_{xx})\right]\sin^4\theta. \quad (12)$$

Figure 4(a) shows the angle dependence $F_M(\theta)$ calculated for $\varepsilon_{zz}=2\times10^{-3}$ at different *B*. With *B* increasing from zero the minimum of the free energy density shifts from $\theta_0=\pi/2$ toward smaller values, and **m** gradually turns toward the field direction as Fig. 4(b) shows. At some magnetic field a second minimum at $\theta=0$ appears, so that $F_M$ has two minima separated by a barrier. With further increasing *B* the first minimum close to $\pi/2$ becomes shallower, while the second minimum becomes deeper. Finally, at $B=B^*$ the first minimum disappears and the magnetization rapidly changes its direction, becoming parallel to **B** [see Fig. 4(b)]. In realistic structures the switching between the two minima occurs at lower *B* values smaller than $B^*$ due to the finite temperature and the presence of fluctuations [37], but in the present analysis we consider that the orientation of **m** corresponds to the first minimum of $F_M$ until $B=B^*$. This corresponds to an experiment at zero temperature with a gradual magnetic field increase starting from zero.

The equilibrium orientation of magnetization determines the response of $\mathbf{B}_{\text{eff}}$ on the strain pulse. As one sees from Eq. (7), at zero shear strain when $\varphi_0 = 0$ the rate $f_\theta = 0$ and the tilt of $\mathbf{B}_{\text{eff}}$ is determined by the value of $f_\varphi$. Figure 5(a) shows the field dependence of the absolute value $|f_\varphi(B)|$ for $\varepsilon_{zz}= 1\times10^{-3}$, $2\times10^{-3}$ and $3\times10^{-3}$. Since $A_{4\varepsilon} \ll A_{2\varepsilon}$ the following approximation can be made: $|f_\varphi(B)| \approx 2\gamma A_{2\varepsilon}\cos\theta_0$, which follows from the field dependence of $m_z(B)$. Therefore $|f_\varphi(B)|$ almost linearly increases with $B$ until the jump at $B=B^*$, as clearly seen from the comparison of Figs. 4(b) and 5(a). The switching field $B^*$ is an increasing function of $\varepsilon_{zz}$ and equals to 117 mT, 180 mT and 243 mT (shown by the vertical dashed lines) for $\varepsilon_{zz}=1\times10^{-3}$, $2\times10^{-3}$ and $3\times10^{-3}$, respectively. Thus, the stronger the magnetization is turned away from the in-plane easy axis by the external magnetic field, the larger is $|f_\varphi|$ and the stronger is the response of $\mathbf{B}_{\text{eff}}$ on the perturbation induced by the strain pulse $\delta\varepsilon_{zz}$.

While $f_\varphi$ determines the tilt of $\mathbf{B}_{\text{eff}}$, the subsequent time evolution of **m** depends significantly on the precession frequency. Fig. 5(b) shows the field dependence of $\omega_0(B)$ for several values of static strain components $\varepsilon_{zz}$. The value of $\omega_0$ decreases with increasing $B$ until it becomes zero at $B=B^*$. The stronger the static epitaxial strain $\varepsilon_{zz}$ is, the larger is $\omega_0$.

The precession rate $f_\varphi$ and the precession frequency $\omega_0$ at a certain external magnetic field are the parameters of the FMS layer, which do not depend on the shape of the strain pulse. However the spatial-temporal evolution of the magnetization is induced by $\delta\varepsilon_{zz}(z,t)$. We calculate the magnetization evolution at three coordinates in the FMS layer: $z=0$, 100 nm and 190 nm. Fig. 3(b) shows the corresponding numerical solution for the component $\delta m_z(t)= \delta\theta(t)\sin\theta_0$. We see that the precession starts upon arrival of the strain pulse at the corresponding coordinate in the FMS layer. While the strain pulse propagates forwards and backwards the precession trajectory is complicated. When the reflected strain pulse completely has left the layer ($t$=110 ps shown by the vertical line) the magnetization continues to precess without decay as long as damping does not occur.

In the considered case of zero shear strain the simple analytical solutions of Eq.(6) for the after-pulse, free magnetization precession can be written as harmonic oscillations with frequency $\omega_0$ which are shifted in phase by $\pi/2$ relative to each other:

$$\delta\varphi(z,t) = 2 f_\varphi S_\omega \sin\left(\frac{\omega_0 \Delta t}{2}\right) \cos \omega_0 (t - t_d),$$
$$\delta\theta(z,t) = -2 a_\perp f_\varphi S_\omega \sin\left(\frac{\omega_0 \Delta t}{2}\right) \sin \omega_0 (t - t_d).$$
(13)

where $\Delta t = 2(d-z)/v_l$ is the travel time of the strain pulse from the coordinate $z$ toward the surface, and back; $t_d = d/v_l$ is the travel time of the strain pulse through the magnetic layer and

$$S_\omega = \varepsilon_{zz}^{\max} \omega_0 \tau^2 \sqrt{2\pi e} \exp(-\omega_0^2 \tau^2 / 2)$$
(14)

is the absolute value of the spectral density of the incident strain pulse at frequency $\omega_0$. For the chosen parameters of the strain pulse $S_\omega$ is an increasing function of frequency in the considered range around $\omega_0$. The parameter $a_\perp = -4 \frac{\gamma}{\omega_0}(B_c - A_{4\varepsilon}(\varepsilon_{zz} - \varepsilon_{xx}))\sin^3 \theta_0$ depends on magnetic field and has values between 0.5 and 1, increasing with increasing magnetic field. The presence of this parameter shows that the precession trajectory of **m** is elliptical with one main axis parallel to the layer plane.

To summarize this part of analysis, the amplitude of precession is determined by three main factors. The first one is the precession rate $f_\varphi$, which describes how sensitive the tilt of effective magnetic field **B**$_{\text{eff}}$ to the strain-pulse induced modulation is. The second one is the spectral density of the incident strain pulse at the precession frequency $\omega_0$. The third one is the oscillating factor $\sin(\omega_0 \Delta t/2)$, which describes the efficiency of interference between incident and reflected parts of the strain pulse at a given coordinate $z$. The maximum amplitude is obtained at a coordinate, where $\Delta t$ is equal to half of the precession period. For $B = 40$ $mT$ and $\varepsilon_{zz} = 2 \times 10^{-3}$ shown in Fig. 3(b), $\omega_0/2\pi = 6.2$ $GHz$, and maximum amplitude is reached at $\Delta t = 80\,ps$ corresponding to z=0. The dependence of the components $\delta m_y = \delta\varphi \cos\theta_0$, which is almost twice larger than $\delta m_z$, and $\delta m_x = \delta\theta \cos\theta_0$, are very similar to $\delta m_z$, and therefore, we do not plot them separately.

We also solve the dynamical equations for the averaged values $\overline{\delta\varphi}(t)$ and $\overline{\delta\theta}(t)$, which are as well harmonic oscillations shifted by $\pi/2$ relative to each other:

$$\overline{\delta\varphi}(t) = 4 f_\varphi S_\omega \frac{\sin^2(\omega_0 t_d / 2)}{\omega_0 t_d} \cos\omega_0(t - t_d),$$
$$\overline{\delta\theta}(t) = -4 a_\perp f_\varphi S_\omega \frac{\sin^2(\omega_0 t_d / 2)}{\omega_0 t_d} \sin\omega_0(t - t_d).$$
(15)

The precession amplitude of the averaged magnetization is also proportional to $f_\varphi$ and $S_\omega$, but depends on the layer thickness through the oscillating factor $\sin^2(\omega_0 \tau_d/2)/\omega_0 \tau_d$ with the first maximum at $\omega_0/2\pi \approx 10$ $GHz$. The thick red line in Figs. 3(a) and 3(b) shows the evolution of the averaged functions $\overline{\delta\varepsilon}(t)$ and $\overline{\delta m_z}(t)$.

Figure 5(c) shows the field dependence of $\overline{\delta m_z}^{\max}(B)$, the amplitude of the after-pulse oscillations $\overline{\delta m_z}(t)$. $\overline{\delta m_z}^{\max}$ was calculated for several values of epitaxial strain $\varepsilon_{zz} = 1 \times 10^{-3}$, $2 \times 10^{-3}$ and $3 \times 10^{-3}$. These dependences reflect the competition between the sensitivity of $\mathbf{B_{eff}}$ to the strain pulse that increases with magnetic field and the response of $\mathbf{m}$ that decreases with B due to the decrease of $\omega_0$. As a result $\overline{\delta m_z}^{\max}(B)$ has a pronounced maximum $\overline{\delta m_z}^{\max} \approx 10^{-3}$ at an optimal intermediate magnetic field. A stronger static epitaxial strain at a given $B$ leads to an increase of both $f_\varphi(B)$ and $\omega_0(B)$ and, thus, the maximum of $\overline{\delta m_z^{max}}(B)$ also shifts to higher magnetic fields. In general, the field dependence of the precession amplitude, as well as its maximum value of $10^{-3}$, is in good agreement with the experimental results [5].

We also numerically analyze the influence of nonzero positive shear strain $\varepsilon_{xy}$. At finite $\varepsilon_{xy}$ the precession rate $f_\theta$ is nonzero even at $B=0$, but it rapidly decreases and becomes negligible with increasing $B$, see Eq. (7). As a result, for almost the whole range of $B$ the response of $\mathbf{B}_{\text{eff}}$ on the strain-induced perturbation is determined mainly by $f_\varphi$ and is not affected substantially by the presence of shear strain. In Figs. 5(b) and 5(c) we see the decrease of the precession frequency and the precession amplitude over the whole range of $B$ in presence of shear strain. The calculated field dependencies $\overline{\delta m_z^{max}}(B)$ for $\varepsilon_{xy} = 2 \times 10^{-4}$ are shown in Fig. 5(c) by the dash-dotted lines.

### B. In-plane magnetic field

If an external magnetic field is applied along the [100] crystallographic direction and the shear strain is zero, $\mathbf{m}$ is oriented along the [100] axis for any value of $B$ and strain-pulse-induced magnetization precession is impossible. Therefore, the presence of shear strain is a key requirement for this geometry. Below we examine the case of nonzero, but small positive $\varepsilon_{xy}$, for which $\mathbf{m}$ is slightly turned in the film plane toward the [1$\bar{1}$0] direction. In this case the free energy density depends only on $\varphi$ and we may simplify expression (4) for $F_M$ to:

$$F_M(\varphi) = \frac{1}{4}\left[B_c - A_{4\varepsilon}(\varepsilon_{zz} - \varepsilon_{xx})\right](3 + \cos 4\varphi) + \frac{1}{2} A_{2xy}\varepsilon_{xy} \sin 2\varphi - B\cos\varphi. \qquad (16)$$

Figure 6(a) shows the dependence $F_M(\varphi)$ for $\varepsilon_{xy} = 2\times 10^{-4}$ for different $B$. At zero magnetic field $F_M$ has the minimum at finite $\varphi_0 < 0$. With increasing $B$ the minimum gradually shifts toward the [100] axis. Figure 6(b) shows the field dependence of the projection $m_x = \cos\varphi_0$ for two values of $\varepsilon_{xy}$. In contrast to the case of perpendicular magnetic field, where we observe a rapid step-like turn of **m** toward the field direction at some threshold, here **m** continuously is rotated with increasing magnetic field.

For this geometry the tilt of **B**$_{\text{eff}}$ is determined only by $f_\theta$, since $f_\varphi = 0$. So the strain pulse $\delta\varepsilon_{zz}(t,z)$ tilts **B**$_{\text{eff}}$ maintaining, however, its in-plane orientation. Figure 7(a) shows the field dependence of $|f_\theta|$ for two values of $\varepsilon_{xy} = 1\times 10^{-4}$ and $2\times 10^{-4}$. The value of $|f_\theta| = -\gamma A_{4\varepsilon} \sin 4\varphi_0$ decreases with increasing $B$ since **m** is tilted closer to the [100] crystallographic direction. The larger $\varepsilon_{xy}$ is the stronger is the response of **B**$_{\text{eff}}$ on $\delta\varepsilon_{zz}$, while the static epitaxial strain $\varepsilon_{zz}$ does not influence the value of $f_\theta$ substantially. One sees that $f_\theta$ is two orders of magnitude smaller than $f_\varphi$ due to the significant difference in the values of the magnetoelastic coefficients $A_{4\varepsilon}$ and $A_{2\varepsilon}$. Obviously, the strain pulse $\delta\varepsilon_{zz}(t,z)$ affects the in-plane orientation of **B**$_{\text{eff}}$ much weaker.

Figure 7(b) shows the field dependencies of the precession frequency $\omega_0(B)$ for several values of $\varepsilon_{zz}$ and $\varepsilon_{xy}$. Contrary to the case of a perpendicular magnetic field, here $\omega_0$ continuously increases with increasing $B$. However, the dependence of $\omega_0$ on the static strain components is the same: $\omega_0$ is larger for stronger epitaxial strain $\varepsilon_{zz}$ and becomes smaller with increasing $\varepsilon_{xy}$.

We also give simple analytical expressions for the after-pulse free precession of **m**:

$$\delta\varphi(z,t) = \frac{2}{a_{\parallel}} f_{\theta} S_{\omega} \sin\left(\frac{\omega_0 \Delta t}{2}\right) \sin \omega_0 (t - t_d),$$
$$\delta\theta(z,t) = 2 f_{\theta} S_{\omega} \sin\left(\frac{\omega_0 \Delta t}{2}\right) \cos \omega_0 (t - t_d).$$
(17)

and for the corresponding averaged values

$$\overline{\delta\varphi}(t) = \frac{4}{a_{\parallel}} f_{\theta} S_{\omega} \frac{\sin^2(\omega_0 \tau_d / 2)}{\omega_0 \tau_d} \sin \omega_0 (t - \tau_d),$$
$$\overline{\delta\theta}(t) = 4 f_{\theta} S_{\omega} \frac{\sin^2(\omega_0 \tau_d / 2)}{\omega_0 \tau_d} \cos \omega_0 (t - \tau_d).$$
(18)

where $a_{\parallel} = -\frac{\gamma}{\omega_0}\left[4(B_c - A_{4\varepsilon}(\varepsilon_{zz} - \varepsilon_{xx}))\cos 4\varphi_0 + 2 A_{2xy}\varepsilon_{xy}\sin 2\varphi_0 - B\cos\varphi_0\right]$ has a value between 0.5 and 1 and increases with external magnetic field.

Figure 7(c) demonstrates the field dependence of $\overline{\delta m_z^{max}}(B)$, which looks similar to the preceding geometry. The main differences are: (i) the smaller value of $\overline{\delta m_z^{max}}(B)$ due to the significantly smaller value of $f_\theta$ and (ii) a different dependence of the precession amplitude on the static strain component $\varepsilon_{zz}$. For this magnetic field direction the shear strain affects the precession rate $f_\theta$ significantly, which increases with increasing $\varepsilon_{xy}$, but changes only slightly the precession frequency. As a result $\overline{\delta m_z^{max}}$ is much larger for stronger shear strain. In contrast, the static epitaxial strain $\varepsilon_{zz}$ does not change the precession rate $f_\theta$ substantially, but $\omega_0$ is still higher for larger $\varepsilon_{zz}$. As a result, at low magnetic fields this leads to an increase of $\overline{\delta m_z^{max}}$ and to a shift of the maximum to lower $B$ with increasing $\varepsilon_{zz}$. At high magnetic fields, however, $\overline{\delta m_z^{max}}$ is reduced for stronger epitaxial strain $\varepsilon_{zz}$. The crossing occurs at a magnetic field [shown by the dashed line in Fig. 7(c)] at which $\omega_0 / 2\pi = 12$ $GHz$, corresponding to the maximum of the function $S_\omega \sin^2(\omega_0 \tau_d/2)/\omega_0 \tau_d$. Thus, at stronger magnetic fields the increase of the precession frequency leads to a decrease of the precession amplitude as seen in Fig. 7(c).

## 5. SUMMARY

To summarize the developed analysis, we have elaborated three important factors that determine the efficiency of the strain pulse-induced magnetization precession. The first one is how strong the distraction of the magnetization direction from equilibrium by the dynamical strain is for a given orientation and strength of the external magnetic field. The distraction is determined by the magneto-crystalline anisotropy of the FMS layer, which depends on a number of parameters, including the holes and magnetic spins concentrations, the lattice temperature, the growth direction, as well as the presence of a shear static deformation. The MCA characterizes the sensitivity of the magnetic system to the strain pulse but does not vary with the specific shape of the pulse.

The second factor arises from the spectral properties of the strain pulse. The cumulative effect of the pulse is the excitation of precession at the frequency of the ferromagnetic resonance. Naturally, the amplitude of precession is proportional to the spectral density of the strain pulse components at this frequency. For the assumed pulse shape, it is determined by the value of $S_\omega$. It is worth to mention here that for typical strain pulses the spectrum is quite broad, being extended up to a few hundred GHz.

Finally, the third factor appears because of the interference of the incident and reflected parts of the strain pulse. As a result, the precession amplitude averaged over the layer thickness is given by the oscillating function of the ratio of the travel time of the strain pulse through the film and the period of the magnetization precession. Thus for a given ferromagnetic resonance frequency it is possible to predict at which film thickness the excitation of precession is most efficient.

The maximal amplitude achieved for perpendicular orientation of the external magnetic field is $10^{-3}$ and depends on the three factors summarized above. Experimentally strain pulses with 10 times larger amplitudes may be injected into the FMS layer. If in addition the pulse duration, the layer thickness and the precession frequency are perfectly adjusted to each other, the maximal estimated amplitude of precession is $5\times10^{-2}$. For in-plane orientation of the magnetic field the effect of the strain pulse is much weaker due to the much smaller anisotropy coefficients. However, in recent experiments on a variety of (Ga,Mn)As layers the precession amplitude was just twice less for this experimental geometry compared to the case of a perpendicular magnetic field [17]. The difference between the experimental observation and the results of our analysis may arise from the uncertainty of the value of $A_{4\varepsilon}$, which is hard to obtain by steady-state measurements or to calculate accurately in the frame of a microscopic model. For a larger value of the cubic magnetoelastic coefficient $A_{4\varepsilon}$ than assumed here we estimate comparable maximal precession amplitudes for the in-plane field geometry and most importantly for the case of zero magnetic field.

Nevertheless, this value is not enough for strain-induced switching of magnetization between the in-plane easy axes. A much stronger effect may be achieved for a shear strain pulse due to the much larger value of the in-plane uniaxial magnetoelastic coefficient $A_{2xy}$. A strain pulse $\delta\varepsilon_{xy}$ of amplitude $4\times10^{-4}$ may rotate $\mathbf{B}_{\text{eff}}$ completely toward the $[1\bar{1}0]$ direction for the chosen FMS layer parameters. In this case the magnetization will precess between the [100] and [010] directions and if the strain pulse is properly shaped precessional switching of the magnetization in analogy to the experiments with pulsed magnetic fields [2] becomes possible. The idea of precessional switching by modulating the MCA of a (Ga,Mn)As layer has been discussed recently [38] and has also been demonstrated [4], although in another material and at lower frequencies. The analysis of the strain-pulse induced magnetization precession for a shear strain pulse may be done in the same way.

To conclude we have carried out a comprehensive analysis of the magnetization precession induced by a strain pulse in a thin FMS layer. We have chosen a strain pulse shape that is typical for ultrafast acoustic experiments, and modeled numerically the strain-pulse-induced spatial-temporal evolution of magnetization. Solution of the Landau-Lifshitz equation in linear approximation has lead to simple analytical expressions for the amplitude of the strain-pulse-induced precession both for any point in the FMS layer as well as averaged over the whole layer. We have found that strain-pulse-induced precession becomes possible when in equilibrium the magnetization is not parallel to the main crystallographic axes and in-plane diagonals. This condition is fulfilled in presence of shear strain in-plane anisotropy or in an external magnetic field, which turns the magnetization out of the easy axis. We have numerically examined two alternative directions of the magnetic field and analyzed the dependence of the precession amplitude on the field strength and the static strain components. The value of epitaxial strain mainly influences the precession frequency and in that way slightly affects the precession amplitude. The shear strain becomes crucially important for in-plane magnetic fields and mainly determines the precession amplitude in this geometry.

ACKNOWLEDGEMENTS

We thank A.V. Akimov, R.V. Pisarev and B. A. Glavin for valuable discussions. The work was supported by the Deutsche Forschungsgemeinschaft (Grant No. BA1549/14-1), the Russian Foundation for Basic Research (11-02-00802), the Russian Academy of Sciences, and the National Science Foundation (grant DMR10-05851).

# APPENDIX A: ANISOTROPY COEFFICIENTS CALCULATION

The anisotropy coefficients $B_c, A_{2\varepsilon}, A_{4\varepsilon}^{(1)}, A_{4\varepsilon}^{(2)}$ and $A_{2xy}$ for a particular structure, which determine the response of the MCA to the strain pulse, can be obtained experimentally, e.g., from ferromagnetic resonance or magneto-transport measurements, or calculated using a microscopic theory. A thorough comparison of experimental and theoretical data may be found in Ref. [29]. Theoretical approaches to the ferromagnetism of (Ga,Mn)As are largely based on the Zener mechanism originally proposed for metals [39,40] and assume that the ferromagnetic coupling between the Mn spins is mediated by free holes [7,8,18]. The free energy density can be calculated using the effective mass Hamiltonian which, in addition to the six-band $\mathbf{k} \cdot \mathbf{p}$ Luttinger-Kohn and the strain terms, includes the $p$-$d$ exchange interaction of the holes and the Mn spins in the molecular-field approximation [18]. According to this model the mechanism of the strain-pulse induced precession is that the pulse changes the hole spectrum, giving rise to a hole redistribution among the energy bands. This, in turn, results in a change of the magnetization orientation according to the minimum of the free energy. Using this model we calculate the intrinsic anisotropy parameters. The hole spectrum calculations are done in the limits of $T=0$ and $B=0$, in accordance to Refs. [9,18,29]. The parameters of the Hamiltonian are chosen like in Ref [18] with the only difference that the shear deformation component is taken into account according to Refs. [41,42]. Since the hydrostatic strain ($\varepsilon_{xx} = \varepsilon_{yy} = \varepsilon_{zz}$) does not affect the magnetic anisotropy in this model the additional relation $A_{4\varepsilon}^{(1)} = A_{4\varepsilon}^{(2)}$ between the magnetoelastic constants is fulfilled.

The $p$-$d$ exchange interaction is described in Ref. [18] by the parameter which is proportional to the number of Mn spins $N_{Mn}$. Since the presence of Mn interstitial defects reduces the number of active Mn spins the real number of Mn spins interacting with the holes is smaller than the one introduced by the nominal doping $x_{Mn} = 0.05$ [7,8]. To account for that, we calculate the anisotropy parameters as function of saturation magnetization $\mu_0 M_0$ for a range of Mn ion concentrations $x_{Mn} = 0.03 \div 0.05$ and for a range of hole concentrations $p = (1 \div 5) \times 10^{20} cm^{-3}$. The best agreement with the experiment reported in Ref. [5] has been obtained for the following parameters: $x_{Mn}$=0.045; $p$=4×10$^{20}$ cm$^{-3}$ and $\mu_0 M_0$=60 mT. The cubic anisotropy field and magnetoelastic coefficients obtained for these parameters are: $B_c = -35$ mT, $A_{2\varepsilon}$=25 T and $A_{2xy}$=152 T. It is difficult to determine reliably the coefficient $A_{4\varepsilon}$ because of its negligibly small value compared to the other coefficients. For this reason it is usually taken as zero [9,18,29]. It follows from the experimental data and the dependence of $B_c$ on the lattice-mismatch strain $\varepsilon_{zz}$ that $A_{4\varepsilon}$ is negative and the value of $A_{4\varepsilon}\varepsilon_{zz}$ is an order of magnitude smaller than $B_c$ [9,26,37]. Therefore we take $A_{4\varepsilon} = -0.5$ T for the calculations.

FIGURE CAPTIONS

**Figure 1.** (a) Schematic of experiments with picosecond acoustic pulses in ferromagnetic epitaxial layers. (b) Temporal profile of the strain pulse injected into the GaAs substrate from the metal film.

**Figure 2.** (a) Equilibrium orientation of the effective field $\mathbf{B}_{\text{eff}}$ and the magnetization $\mathbf{M}$ in perpendicular external magnetic field $\mathbf{B}$, and coordinate system orientations used in the article. (b) Magnetization precession after the strain pulse has left the FMS layer.

**Figure 3.** (a) Temporal evolution of the strain pulse $\delta\varepsilon_{zz}(t,z)$ at three positions in the FMS layer (black lines) and the relative modulation of the layer thickness $\overline{\delta\varepsilon}(t)$ (thick red line).
(b) Strain-pulse-induced temporal evolutions of the magnetization projection $\delta m_z(t,z)$ at three positions in the FMS layer (thin black lines) and the value averaged across the layer $\overline{\delta m_z}(t)$ (thick red lines). The evolutions are calculated at $B$=40 mT applied perpendicular to the layer plane under static strain $\varepsilon_{zz}=2\times10^{-3}$ and $\varepsilon_{xy}=2\times10^{-4}$. Time $t=0$ corresponds to the moment when the center of the incident strain pulse crosses the GaAs /(Ga,Mn)As interface. The vertical dot-dashed line shows the time moment at which the strain pulse leaves the FMS layer.

**Figure 4.** (a) Normalized free energy density $\Delta F_M = F_M(\mathbf{m}) - F_M(m_x)$ as function of angle $\theta$ for different values of the external magnetic field $B$ applied perpendicular to the layer. (b) Field dependence of the magnetization projection $m_z = \cos\theta_0$ onto the direction of magnetic field for three values of the static epitaxial strain. The vertical dashed lines show the values of $B^*$ when $\mathbf{m}$ rapidly turns toward the external field direction (see text). The calculations are done for $\varepsilon_{xy}=0$.

**Figure 5**. Magnetic field dependencies of the absolute value of the effective precession rate $|f_\varphi|$ (a), the precession frequency $\omega_0/2\pi$ (b), and the averaged precession amplitude $\overline{\delta m_z^{\max}}$ (c) for $\mathbf{B}$ perpendicular to the layer plane, calculated for different values of the static strain components $\varepsilon_{zz}$ and $\varepsilon_{xy}$.

**Figure 6.** (a) Normalized free energy density $\Delta F_M = F_M(\mathbf{m}) - F_M(m_x)$ as function of the equilibrium angle $\varphi$ for different values of $B$ applied along the [100] direction in presence of shear strain. (b) Field dependence of the magnetization projection $m_x = \cos\varphi_0$ onto the direction of the magnetic field for two values of shear strain $\varepsilon_{xy}$.

**Figure 7.** Magnetic field dependencies of the absolute value of the effective precession rate $|f_\theta|$ (a), the precession frequency $\omega_0/2\pi$ (b) and the averaged precession amplitude $\overline{\delta m_z^{\max}}$ (c) for $\mathbf{B} \parallel [100]$ calculated for different values of the static strain components $\varepsilon_{zz}$ and $\varepsilon_{xy}$. The dashed lines show the frequency (horizontal) and the corresponding value of magnetic field (vertical) demarking the field-frequency range in which a higher precession frequency results in a larger precession amplitude.

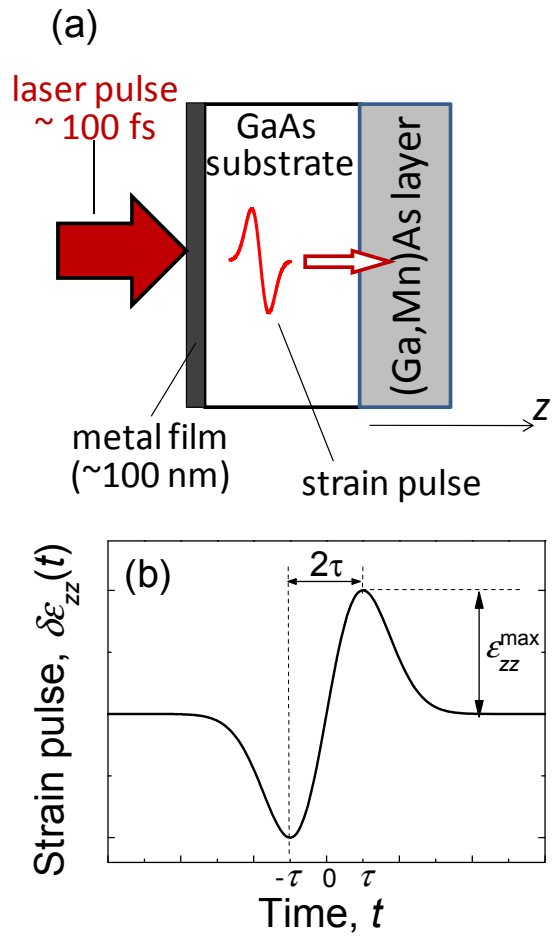

**Figure 1.** *T. L. Linnik et al. "Theory of magnetization precession…"*

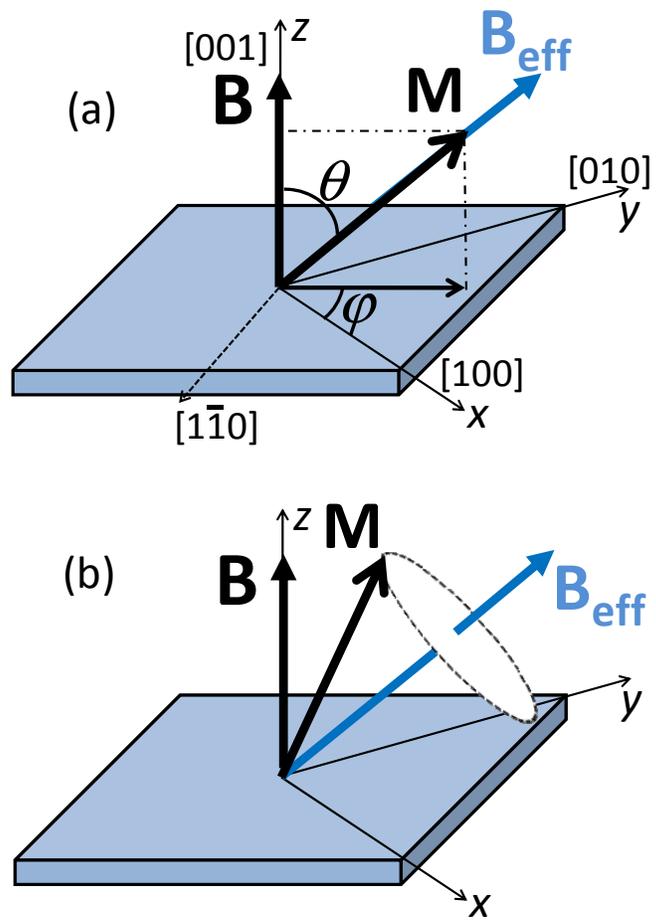

**Figure 2.** *T. L. Linnik et al. "Theory of magnetization precession…"*.

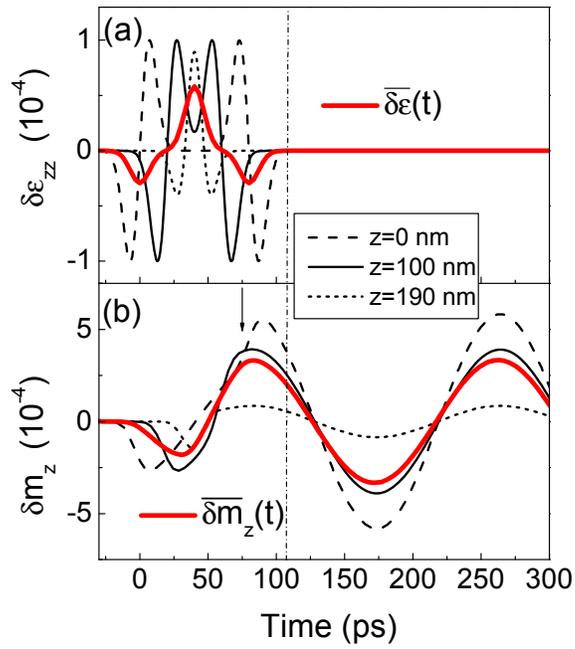

**Figure 3.** *T. L. Linnik et al. "Theory of magnetization precession…"*

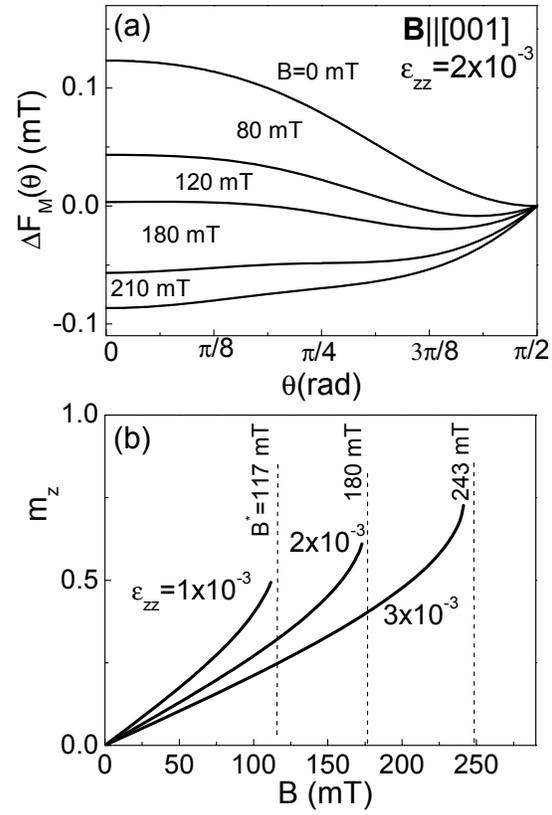

**Figure 4.** *T. L. Linnik et al. "Theory of magnetization precession…"*

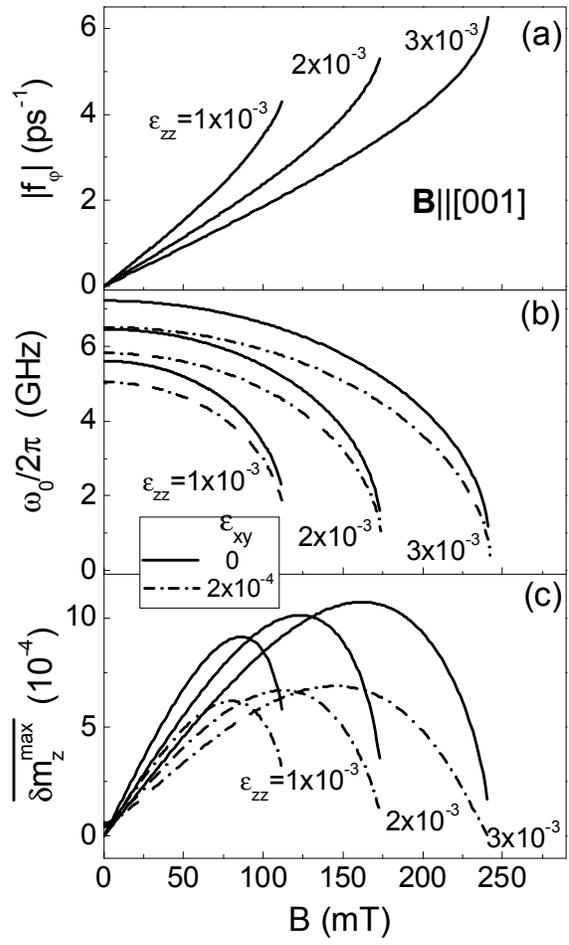

**Figure 5**. *T. L. Linnik et al. "Theory of magnetization precession…"*

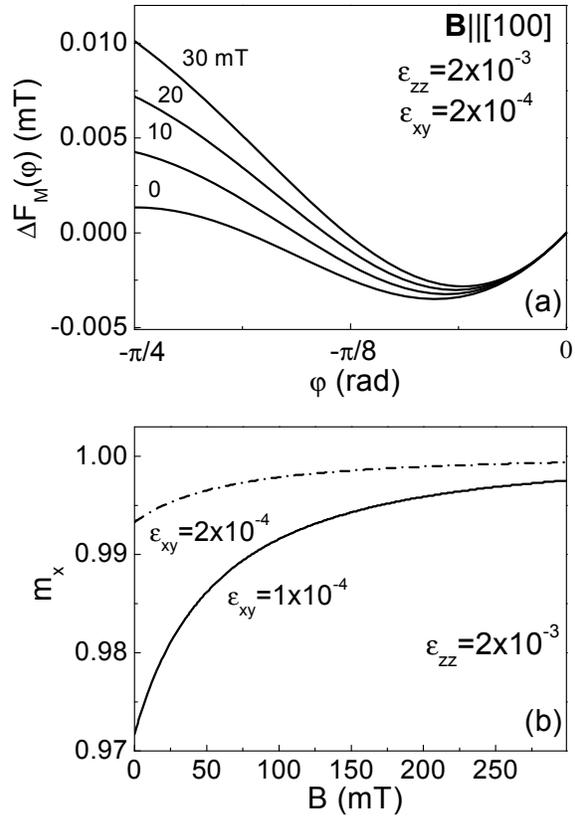

**Figure 6.** *T. L. Linnik et al. "Theory of magnetization precession…"*

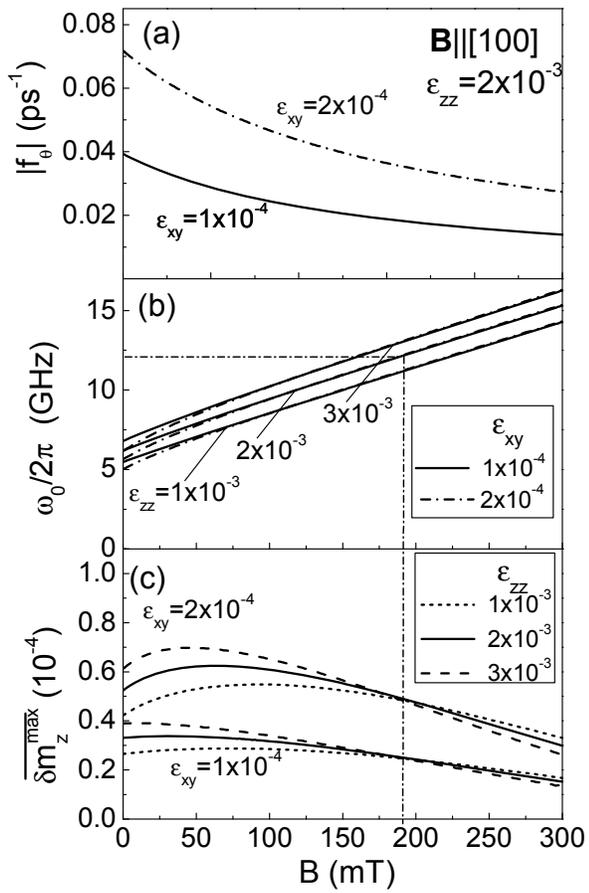

**Figure 7.** *T. L. Linnik et al. "Theory of magnetization precession…"*